\documentclass[12pt,floats]{iopart}
\usepackage{graphicx}

\def\lsim{\mathrel{\rlap{\lower4pt\hbox{\hskip1pt$\sim$}}
    \raise1pt\hbox{$<$}}} 
\def\gsim{\mathrel{\rlap{\lower4pt\hbox{\hskip1pt$\sim$}}
    \raise1pt\hbox{$>$}}}  
\makeatother
\begin{document}

\title{Improving the Sensitivity of LISA}

\author{K.~Rajesh~Nayak\dag, A.~Pai\ddag, S.~V.~Dhurandhar\dag~ and
  J-Y.~Vinet\ddag}
\address{\dag Inter-University Centre for Astronomy and Astrophysics, Pune, 
India.}
\address{\ddag ILGA, Dept. Fresnel, Observatoire de la Cote d'Azur, 
 BP 4229, 06304 Nice, France.}

\ead{nayak@iucaa.ernet.in, archana@obs-nice.fr,sanjeev@iucaa.ernet.in, vinet@obs-nice.fr}

\begin{abstract}
It has been shown in several recent papers that the six Doppler data streams obtained 
from a triangular LISA configuration can be combined by appropriately delaying 
the data streams for cancelling the laser frequency noise. Raw laser noise 
is several orders of magnitude above the other noises and thus it is essential to bring it
down to the level of other noises such as shot, acceleration, etc. A rigorous and 
systematic formalism using the powerful techniques of computational commutative algebra was 
developed which generates in principle {\it all} the data combinations cancelling the 
laser frequency noise. The relevant data combinations form a first module of syzygies. 
\par
In this paper we use this formalism to advantage for optimising the sensitivity of LISA 
by analysing the noise and signal covariance matrices. The signal covariance matrix is 
calculated for binaries whose frequency changes at most adiabatically and the signal is 
averaged over polarisations and directions. We then present the extremal SNR curves for 
all the data combinations in the module. They correspond to the eigenvectors of the 
noise and signal covariance matrices. A LISA `network' SNR is also computed by combining 
the outputs of the eigenvectors. We show that substantial gains in sensitivity can be 
obtained by employing these strategies. The maximum SNR curve can yield an improvement 
upto 70 \% over the Michelson, mainly at high frequencies, while the improvement using 
the network SNR ranges from 40 \% to over 100 \%. 
\par 
Finally, we describe a simple toy model, in which LISA rotates in a
plane. In this analysis, we estimate the improvement in
the LISA sensitivity, if one switches from one data combination to 
another as it rotates. Here the improvement in sensitivity,
if one switches optimally over three 
cyclic data combinations of the eigenvector is about 55 \% on an average over the LISA 
band-width. The corresponding SNR improvement increases to 60 \%, if one maximises over 
the module. 
 
\end{abstract}
\maketitle
\section{Introduction}

The proposed space-based mission of gravitational wave (GW) detection - the 
Laser Interferometric Space Antenna (LISA) - consists of three identical
spacecrafts forming an equilateral triangle of side $5\times 10^{6}$ km 
following heliocentric orbits. LISA is thus a giant interferometric configuration with
three arms which will give independent information on GW polarisations.
The implementation of this huge Michelson interferometer is quite 
different from the ground based interferometers like LIGO, VIRGO etc.
Unlike in the ground based interferometric detectors, it is not feasible 
to bounce the laser beams because of the large armlengths involved. A special
Doppler tracking scheme is used to track the space crafts with laser
beams. This exchange of laser beams between the three space-craft 
result in six Doppler data streams.  

LISA sensitivity is limited by several noise sources. A major noise source is the   
laser phase noise which arises due to phase fluctuations of the master laser.  
Amongst the important noise sources, laser phase noise is expected
to be several orders of magnitude larger 
than other noises in the instrument. The current  
stabilisation schemes estimate (this estimate may improve in the future) this noise to about 
$\Delta \nu /\nu_0 \simeq 10^{-13}/\sqrt{{Hz}}$, where $\nu_0$ is the frequency of the 
laser and $\Delta \nu$ the fluctuation in frequency.
If the laser frequency noise can be suppressed then the noise floor is determined by the  
optical-path noise which fakes fluctuations in the lengths of optical
paths and the residual acceleration of proof masses resulting from imperfect 
shielding of the drag-free system. The noise floor is then at an effective GW strain 
sensitivity $h \sim 10^{-21}$ or $10^{-22}$. 
Thus, cancelling the laser frequency noise is vital if LISA is to reach the requisite 
sensitivity. Since it is impossible to maintain equal distances between space-craft, 
cancellation of laser frequency noise is a non-trivial problem.  
Several schemes have been proposed to combat this noise. In these schemes \cite{AET99,ETA00}, 
the data streams are combined with appropriate time delays in order to cancel the 
laser frequency noise. In our earlier work \cite{SNV02}, henceforth referred to as paper I, 
we had presented a systematic and rigorous method using commutative algebra which 
generates {\it all} 
the data combinations cancelling the laser frequency noise. These data combinations 
form a module over a polynomial ring, well known in the literature, as 
the \emph{first module of syzygies}. We obtained the generators of this module and hence
the entire set of relevant data combinations could be generated conveniently. More 
importantly, we note that this method is general and can be extended to 
space-missions with more than three spacecrafts.

In this paper we employ our previously set up formalism for two important applications:
We compute the noise covariance matrix for LISA and obtain its eigenvectors and eigenvalues.
The signal covariance matrix is computed for binaries whose frequency changes at most 
adiabatically (the monochromatic case is included) and for which the signal is averaged 
over polarisations and directions. Here adiabatic means that the signal response, the 
noise and hence the SNR change imperceptibly even if the GW source changes frequency 
during the observation time. Thus, even though the results are presented at each fixed 
frequency, the sources need not be strictly monochromatic, and apply to a wider class 
of sources. The signal covariance matrix has the same eigenvectors as the noise covariance 
matrix which results in computational simplification. We show that 
the signal-to-noise (SNR) for any data combination in the module, then lies between  
an upper and a lower bound. The upper and lower bounds of the SNR are functions of 
frequency which are just the SNR curves of the eigenvectors. 
The extremisation - both maximisation and minimisation 
- of SNR is important for different purposes; maximisation is important for the detection 
and parameter estimation of a GW source, while 
minimisation is important for the purpose of distinguishing the GW confusion noise from 
the instrumental noise \cite{CNFSN}. We further show that the bounding SNR curves have multiple 
intersections within the band-width of LISA $(10^{-4} - 1)$ Hz. 
The improvement of SNR of the upper-bound over the Michelson combination goes upto 70 \% , 
but only at high frequencies $\gsim 5$ mHz. At low frequencies $\lsim$ 5 mHz, both 
have the same sensitivity. Since the eigenvectors are 
independent random variables, a `network' SNR  of independent detectors \cite{SDP99} 
can be constructed from the likelihood 
considerations which gives an improvement between $\sqrt {2}$ and $\sqrt {3}$ over the 
maximum of SNRs of the eigenvectors. The improvement over the Michelson combination is 
about 40 \% at low frequencies $\lsim 3$ mHz and rises above 100 \% at high frequencies. 
We may note that, some of our results are in agreement with
independent and simultaneous calculations by Prince {\it et al} \cite{TTLA02}.

Tracking a GW source fixed on the celestial sphere for determining the SNR constitutes 
a non-trivial problem, which will require substantial analysis when writing codes. Here, 
we consider a simple toy model of LISA rotating in its own plane. Our goal is 
to obtain a rough estimate in the improvement of SNR by optimally switching  
from one data combination to another. We show that, if we implement this strategy, 
it is possible to improve the SNR by about 55 \%.

\section{The module of syzygies of time delayed data combinations}

\begin{figure}
\begin{center}
\caption{\label{cap:Lisa-geometry}The LISA constellation consist of three spacecraft 
each carrying two optical bench systems $i$ and $i^{*}, i = 1, 2, 3$. They exchange 
six laser beams which are represented by $U^{i}$and $V^{i}$ in the
figure. The noise cancellation data combinations are obtained by appropriately
delaying these six beams across the length of arms $L_{i}$. }

\includegraphics[  width=8cm,
  height=8cm]{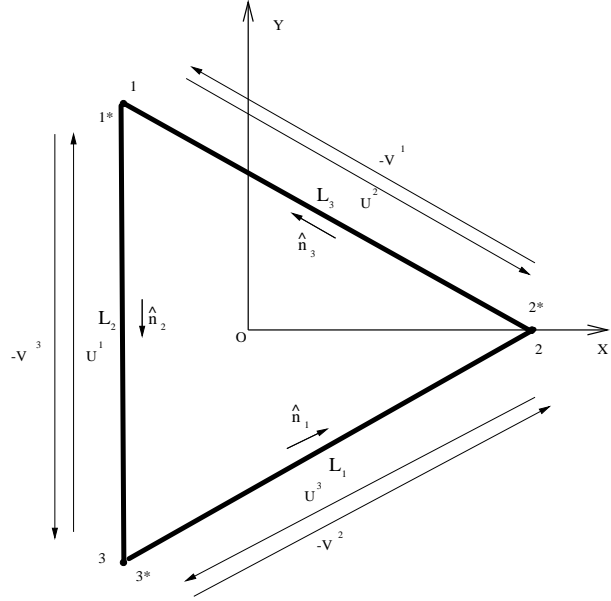}\\
\end{center}
\end{figure}

In this section, we briefly summarise the main results of paper I. 
It has been shown earlier in the literature 
\cite{AET99,ETA00,SNV02}, that the laser phase noise and the optical
bench motion noise in the LISA can be suppressed by combining
the six beams appropriately delayed across arms of the interferometer. 
The six data streams are labelled as $U^i$ and $V^i$, $i = 1, 2, 3$.  
The geometry of the LISA along with six laser beams is shown in the
Figure -\ref{cap:Lisa-geometry}. The beams $U^i$ go clockwise while the $V^i$ go 
counter-clockwise. There are also six other beams connecting the 
two optical benches on each space-craft (which are not shown in the
figure). In the SNR analysis only the {\it difference} between these
beams is important. These differences are denoted 
by $Z^i$ where the index $i$ corresponds to each space-craft labelled $i$.

In paper I, we have shown that \emph{all} the data combinations
which cancel the laser frequency noise and the bench motion noise form a module 
over a ring of polynomials
in the three time delay operators $E_{i}$, where $E_i, i = 1, 2, 3$ 
represent delay operators of the light travel time along the three arms with 
lengths $L_i$. Thus for the data stream $ x(t)\,:\,E_i x (t) = x (t - L_i)$ (the speed of 
light is set to unity); the operator $E_i$ delays the data stream by amount $L_i$. 
For the LISA configuration $L_i \sim 16.7$ seconds, corresponding to an arm-length 
of 5 million km. An important advantage of this formalism is that, one can generate the
entire module from the generators; linear combinations of the generators 
with polynomial coefficients in the ring generate this noise cancellation module. Several sets of 
generators have been listed in paper I.
Based on the physical considerations or for the purpose 
of computational convenience, we may choose one set 
over another. For the purpose of extremising the SNR, we choose the set of generators:
$\alpha, \beta, \gamma$ and $\zeta$ (notation followed from \cite{AET99,ETA00,SNV02}). 
The $\alpha, \beta, \gamma$ are cyclic permutations of each other. This symmetry comes in  
useful when computing scalar products between them and also in 
diagonalising the noise, and signal covariance matrices defined in the next section. 
Following paper I, we list this generating set as 9-tuples of polynomials 
$\left(p_{i},\, q_{i},\, r_{i}\right)$. The polynomials 
$\left(p_{i},\, q_{i},\, r_{i}\right)$ are polynomials in the variables $E_i$ and 
act on the data streams $\left(V^{i},\, U^{i},\, Z^{i}\right)$ respectively. 
The linear combination $p_i V^i + q_i U^i + r_i Z^i$ cancels the laser frequency noise 
and optical bench motion noise when the 9-tuple $\left(p_{i},\, q_{i},\, r_{i}\right)$ 
is an element of the module of syzygies. The generators of the module are given by:   
{\small
\begin{eqnarray}
\hspace*{-1cm} \alpha  =  (1,E_{3},E_{1}E_{3},1,E_{1}E_{2},E_{2},-(1+E_{1}E_{2}E_{3}),-(E_{1}E_{2}+E_{3}),-(E_{1}E_{3}+E_{2}))\, ,\nonumber \\
\hspace*{-1cm} \beta  =  (E_{1}E_{2},1,E_{1},E_{3},1,E_{2}E_{3},-(E_{1}E_{2}+E_{3}),-(1+E_{1}E_{2}E_{3}),-(E_{1}+E_{2}E_{3}))\, ,\nonumber \\
\hspace*{-1cm} \gamma   =  (E_{2},E_{2}E_{3},1,E_{3}E_{1},E_{1},1,-(E_{2}+E_{1}E_{3}),-(E_{1}+E_{2}E_{3}),-(1+E_{1}E_{2}E_{3}))\, ,\nonumber \\
\hspace*{-1cm} \zeta  =  (E_{1},E_{2},E_{3},E_{1},E_{2},E_{3},-(E_{1}+E_{2}E_{3}),-(E_{2}+E_{1}E_{3}),-(E_{3}+E_{1}E_{2}))\, .\label{eq:GEN4}
\end{eqnarray}
}

 We exhibit another set of generators $\left.\right\{ X^{(A)}\left\} \right.$ which 
was obtained from the software package Macaulay 2 \cite{M2}. They can be related to the above 
set of generators by:

\begin{eqnarray}
X^{(1)} & = & E_{3}\, \zeta -\gamma \, ,\nonumber \\
X^{(2)} & = & \zeta \, ,\nonumber \\
X^{(3)} & = & \alpha \, ,\nonumber \\
X^{(4)} & = & \beta \, .\label{eq:RLT}
\end{eqnarray}

 Note that these sets of generators are not linearly independent.
In particular, the set of generators $\{\alpha ,\beta ,\gamma ,\zeta
\}$ \cite{AET99} obey the following condition :
\begin{equation}
(1-E_{1}E_{2}E_{3})\zeta =(E_{1}-E_{2}E_{3})\alpha +(E_{2}-E_{1}E_{3})\beta 
+(E_{3}-E_{1}E_{2})\gamma \, .
\label{eq:CONST}
\end{equation}
When maximising the SNR in the Fourier space, this relation allows us to eliminate one 
of the generators at almost all frequencies, except when the product $E_1 E_2 E_3 = 1$.
Note that $E_1 E_2 E_3$ is just the total time-delay $L_1 + L_2 + L_3$ around the  
LISA triangle. For the purposes of this paper, we assume that all the arms of LISA are 
of equal length \emph{i.e} $L_{1}=L_{2}=L_{3}=L$. In the Fourier domain, 
\emph{i.e, $E_{1}=E_{2}=E_{3}=E=e^{i\Omega L}$} and the operator polynomials become 
actual polynomials. One can then solve 
for $\zeta$ in terms of $\alpha, \beta, \gamma$,  
{\it except} at the frequencies $\Omega$, when $e^{3i\Omega L} = 1$. 
Taking $L \sim 5 \times 10^5$ km, the smallest such frequency 
$f = \Omega / 2 \pi $ is $ \sim 20 $ mHz. Thus, 
\begin{equation}
\zeta =\frac{E}{1+E+E^{2}}\left(\alpha +\beta +\gamma \right)\, ,
\label{eq:RLTaprox}
\end{equation}
and can be effectively eliminated while extremising SNR, except at the singular 
frequencies. Since SNR computation  
can be successfully carried out for frequencies arbitrarily close to the singular 
frequencies, for the computation, the singularities do not seem to be important.
In the analysis that follows, we use only the three generators $\{\alpha ,\, \beta ,\, \gamma \}$.

\section{Strategies for improving the effective sensitivity of LISA}

In this section, we show that the set of generators $\{\alpha ,\, \beta ,\, \gamma \}$
can be combined into a new set, consisting of `orthogonal' eigenvectors. The 
noise covariance matrix naturally defines a positive definite, non-degenerate bilinear 
form, which serves as a scalar product or a metric. Orthogonality between eigenvectors is 
defined in terms of this metric. Physically this means that the noises of the  
eigenvectors are uncorrelated with each other. The eigenvectors are easily obtained 
by diagonalising the noise covariance matrix. The averaged signal 
matrix that we consider here has the same form as the noise covariance matrix and 
consequently has the same eigenvectors. Thus this 
set of eigenvectors simultaneously diagonalises both matrices constituting the 
SNR and simplifies the analysis that follows. An important observation here is that 
the eigenvectors are independent observables. They represent therefore
statistically independent detectors (so far as instrumental noises are
concerned), and they can be treated as a network of detectors.
 Furthermore, they can be combined in a root mean square 
fashion to yield a `detector network statistic' \cite{SDP99} to yield a much improved 
sensitivity.

\subsection{The noise covariance matrix}

Following the formalism in paper I, we define noise vectors in the Fourier domain 
$N^{(I)}, ~I = 1, 2, 3$ for each of the generators $\{\alpha ,\beta ,\gamma \}$ 
respectively, over the 12 dimensional complex space $\mathcal{C}^{12}$,
\begin{equation}
N^{(I)}\, \, =\, \, \left(\sqrt{S^{pf}}(2p_{i}^{(I)}+r_{i}^{(I)}),
\sqrt{S^{pf}}(2q_{i}^{(I)}+r_{i}^{(I)}),\sqrt{S^{sh}}p_{i}^{(I)},
\sqrt{S^{sh}}q_{i}^{(I)}\right)\, ,
\label{eq:NOISEVEC}
\end{equation}
 where $S^{pf}(f)$ and $S^{opt}(f)$ are power spectral
densities (psd) of the proof mass residual motion and the optical path noise
respectively. The polynomials $(p_{i}^{(I)}, q_{i}^{(I)}, r_{i}^{(I)})$, 
(now actual polynomials in the Fourier domain) corresponding to the generators 
$\alpha $, $\beta $ and $\gamma $ are given in
the equation (\ref{eq:GEN4}). We take $S^{pf}(f)=2.5\times 10^{-48}\left[f/1\: Hz\right]^{-2}\, Hz^{-1}$
and $S^{opt}(f)= 1.8\times 10^{-37}\left[f/1\: Hz\right]^{2}\, Hz^{-1}$
following the literature \cite{ETA00}. It can be easily shown that
for a given data combination, the norm of the noise vector represents
its noise psd. The noise covariance matrix for the generators 
 $\{\alpha ,\beta ,\gamma \}$ is defined as $\mathcal{N}^{(IJ)}=N^{(I)}\cdot N^{(J)*}$
and takes the simple form, 
\begin{equation}
\mathcal{N}^{(IJ)}=\left[\begin{array}{ccc}
 n_{d} & n_{o} & n_{o}\\
 n_{o} & n_{d} & n_{o}\\
 n_{o} & n_{o} & n_{d}\end{array}
\right].
\label{eq:Nomat}
\end{equation}
We note that because of the cyclic symmetry, the diagonal elements 
$N^{(I)}\cdot N^{(I)*}$ are equal to each other - denoted by  
$n_d$. Similarly, all the off-diagonal elements $N^{(I)}\cdot N^{(J)*}$,
for $I\neq J$ are also equal to each other and which we denote by $n_o$. 
This was the reason a generating set possessing symmetry properties was chosen in 
the first place. A matrix with this form has two degenerate eigenvalues.  
Thus, the eigenvalues of the noise covariance matrix are given by, 
\begin{equation}
n_{1}=n_{2}=n_{d}-n_{o}\; \hspace {1cm}{\mathrm{and}}\hspace {1cm}\; n_{3}=n_{d}+2n_{o}\, .
\label{eq:nsegn}
\end{equation}
 Since two of the eigenvalues are degenerate we need to systematically adopt
a procedure for choosing the linearly independent and orthonormal
set of eigenvectors. This choice is not unique. One such choice gives
the following matrix ${\cal M}$ 
\begin{eqnarray}
\mathcal{M} & = & \left[\begin{array}{ccc}
 \frac{1}{\sqrt{6}} & \frac{1}{\sqrt{6}} & -\sqrt{\frac{2}{3}}\\
 -\frac{1}{\sqrt{2}} & \frac{1}{\sqrt{2}} & 0\\
 \frac{1}{\sqrt{3}} & \frac{1}{\sqrt{3}} & \frac{1}{\sqrt{3}}\end{array}
\right]\, ,
\label{eq:Eigenmatrix}
\end{eqnarray}
which diagonalises the noise matrix $\mathcal{N}^{(IJ)}$, that is 
$\mathcal{M}\, \cdot \, \mathcal{N}\, \cdot \, \mathcal{M}^{-1}$ is diagonal, with eigenvalues
as diagonal elements.  The eigenvectors are:
\begin{eqnarray}
Y^{(1)} & = & \frac{1}{\sqrt{6}}\left(\alpha +\beta -2\gamma \right)\, ,\nonumber \\
Y^{(2)} & = & \frac{1}{\sqrt{2}}\left(\beta -\alpha \right)\, ,\nonumber \\
Y^{(3)} & = & \frac{1}{\sqrt{3}}\left(\alpha +\beta +\gamma \right)\, .\label{eq:3basis}
\end{eqnarray}
\begin{figure}
\begin{center}
\caption{\label{cap:Noise}Noise Spectra of combinations $Y^{(I)}$}

\includegraphics{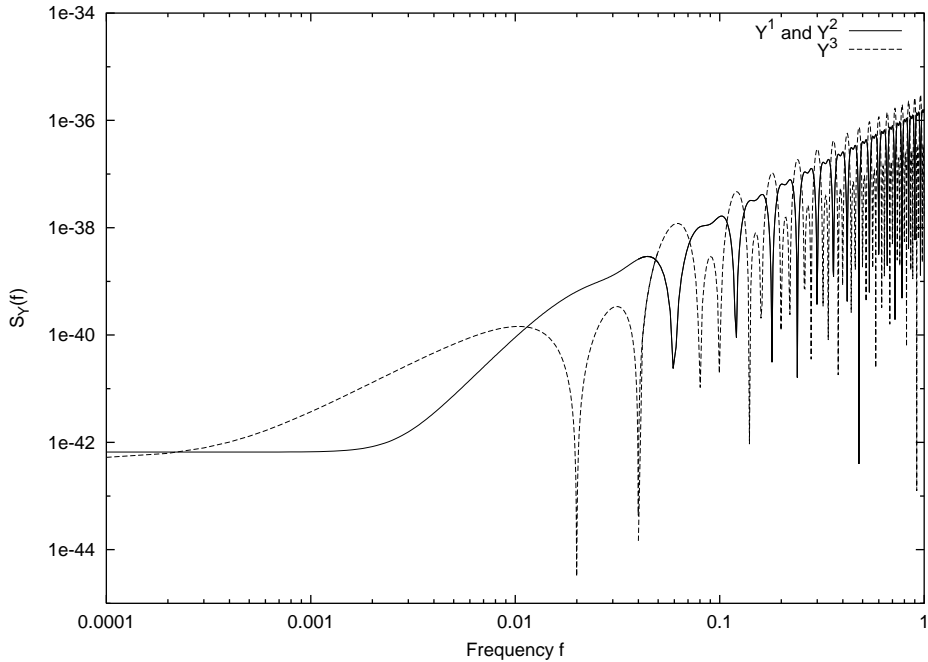}
\end{center}
\end{figure}

We find that the data combination $Y^{(3)}$ is proportional to the symmetric Sagnac
combination $\zeta$ and has the same SNR as that of $\zeta $.

\subsection{The signal covariance matrix}

The response of a GW signal for a given data combination
is computed in paper I. The response is conveniently expressed in 
the Fourier domain and is given by, 
\begin{equation}
h (\Omega ) = \sum _{i=1}^{3}\left[p_{i} \left(F_{V_{i;+}}h_{+}+F_{V_{i;\times }}h_{\times }\right)
+q_{i} \left(F_{U_{i;+}}h_{+}+F_{U_{i;\times }}h_{\times }\right)\right] (\Omega )\, .
\label{eq:gwresp}
\end{equation}
 Here, $F_{V_{i;+/\times }}$and $F_{U_{i;+/\times }}$ are the antenna
pattern functions. 
We note that the signal depends only on the first six entries $(p_i, q_i)$ of the 9-tuple 
describing a data combination. So while dealing with the signal response, we only 
consider the 6-tuple $P = (p_i, q_i)$ of the 
9-tuple describing a data combination.     
We apply this formalism to a binary source which may be adiabatically changing in 
frequency. The two GW amplitudes of such a source at frequency $\Omega$ are given by, 
\begin{eqnarray*}
h_{+}(\Omega ) & = & {\cal A} \left(\frac{1+\cos ^{2}\epsilon }
{2}\cos 2\psi -i\cos \epsilon \, sin2\psi \right) , \\
h_{\times }(\Omega ) & = & {\cal A} \left(-\frac{1+\cos ^{2}\epsilon }
{2}sin2\psi -i\cos \epsilon \, \cos 2\psi \right) .
\end{eqnarray*}
Here, the parameters $\epsilon $ and $\psi $ describe the orientation of the source and 
enter into the expressions for the polarisation amplitudes. The direction of the source 
on the celestial sphere is given by the angles $\theta $ and $\phi $. In order to 
organise the calculations, we also define the detector response 6-tuple as, 
\begin{equation}
R\, =\, \left(F_{V_{i;+}}h_{+}\, +\, F_{V_{i;\times }}h_{\times },\; 
F_{U_{i;+}}h_{+}\, +\, F_{U_{i;\times }}h_{\times }\right)\, .
\label{eq:Fvec}
\end{equation}
Both $P$ and $R$ will be considered as row vectors for the purposes of defining 
matrix products. In order to analyse the signal covariance matrix, it is useful to define 
a scalar product. For two data combinations $P$ and $Q$ (considered as row vectors), 
we define the scalar product as follows:
\begin{equation}
\ll P , Q \gg \, =\, P \cdot \mathcal{R}\cdot Q{\dagger}\, ,
\label{eq:sclar}
\end{equation}
where, $\mathcal{R}\, =\, R^{\dagger} \cdot R $ is a Hermitian matrix of detector  
responses and the `dot' denotes the matrix product. 
The norm of the vector $P$ is then given by, 
\begin{equation}
\left\Vert P \right\Vert^2 \, =\, \ll P , P \gg .
\label{eq:norm}
\end{equation}
The norm of $P$ is the GW response for the data combination described by $P$. 

The signal covariance matrix for any generating set $X^{(I)}$ 
(and corresponding $P^{(I)}$) is then defined as,
\begin{equation}
\mathcal{H}^{(IJ)}=\langle h^{(I)} h^{(J)*}\rangle _{\epsilon \psi \theta \phi }\, 
=\, \left\langle \ll P^{(I)} , P^{(J)}\gg \right\rangle _{\epsilon \psi \theta \phi } ,
\label{eq:sigmat}
\end{equation}
where, $h^{(I)}= P^{(I)} \cdot R^{\dagger}$ and the bracket 
$\langle \; \rangle _{\epsilon \psi \theta \phi }$
represents the average over the polarisations and directions. Taking 
$\alpha $, $\beta $ and $\gamma $ as the generators, the cyclic symmetry 
between them gives rise to a signal covariance matrix $\mathcal{H}^{(IJ)}$ 
which has the same form as the noise covariance matrix  $\mathcal{N}^{(IJ)}$ given 
in equation (\ref{eq:Nomat}). In this case the $n_{d}$ and $n_{o}$ are replaced
by $h_{d}=\langle h^{(I)} h^{(I)*}\rangle $
and $h_{o}=\langle h^{(I)} h^{(J)*}\rangle $ respectively. Thus $\mathcal{H}$ is 
diagonalised by the similarity transformation $\mathcal{M}$ and has the {\it same} 
eigenvectors $Y^{(I)}$. The eigenvalues of $\mathcal{H}^{(IJ)}$ are given by 
\begin{equation}
h_{1}=h_{2}=h_{d}-h_{o}\qquad {\mathrm{a}nd}\qquad h_{3}\, =h_{d}+2h_{o}\, .
\label{eq:sigeign}
\end{equation}

This simultaneous diagonalisation of both signal and noise matrices
is important from the point of extremisation of SNR. This forms the content of the next 
subsection. However, we may note that in the formalism developed by
Prince {\it et al} \cite{TTLA02}, the optimisation is performed
without averaging over the source directions and polarisations, which
results in the GW source matrix of rank $1$. Since, the source
directions are not known in general, we average over these parameters which
results in a signal matrix of rank $3$. 

\subsection{Extremisation of SNR}

An arbitrary data combination can be written as $Y=\alpha _{(I)}Y^{(I)}$,  
where $\alpha_{(I)}$ are polynomials in $E$. The SNR for this combination is given by,  
\begin{equation}
SNR^{2}\, =\, \frac{{k_{1}h_{1}+k_{3}h_{3}}}{{k_{1}n_{1}+k_{3}n_{3}}}\, ,
\label{eq:snr}
\end{equation}
where, $k_{1}= |\alpha _{1}|^{2}+ |\alpha _{2}|^{2}$
and $k_{3}= |\alpha _{3}|^{2}$, and the ranges of $k_1, k_3$ are from 0 to $\infty $.
The simplicity of this expression is because of using the new set of the 
orthogonal generators $Y^{(I)}$. It enables us to easily solve the extremisation 
problem. 
\begin{figure}
\begin{center}
\caption{\label{cap:Plotsens}Log Log plot of sensitivity $S$, curve as function
of $f$ after averaging over polarisation and source directions for
a observation period of one year and SNR =5.}

\includegraphics{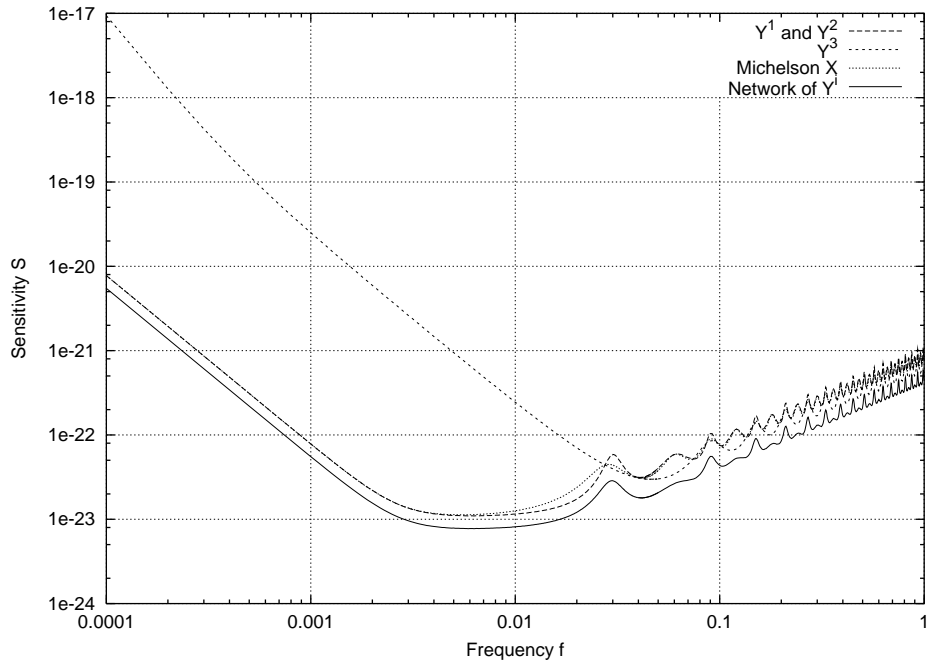}
\end{center}
\end{figure}
The maximum and the minimum of this expression is governed by the quantity $\eta$ where, 
\begin{equation}
\eta (f) =\frac{SNR_{(1)}^{2}}{SNR_{(3)}^{2}}=\frac{h_{1} / n_{1}}{h_{3} / n_{3}}\, ,
\label{eq:ratio}
\end{equation}  
where $SNR_{(I)}$ denotes the SNR of $Y^{(I)}$.
When $\eta (f) \neq 1$, the SNR is a strictly monotonic function of the ratio 
$k_{1} / k_{3}$. 
If $\eta (f) > 1$ then the SNR of the generator $Y^{(1)}$ or $Y^{(2)}$ is greater than 
that of $Y^{(3)}$. Then for these frequencies $Y^{(1)}$ or $Y^{(2)}$ 
(or any linear combination 
of them) yields the maximum SNR while $Y^{(3)}$ yields minimum SNR. When  $\eta (f) < 1$, 
the opposite is true: $Y^{(1)}$ or $Y^{(2)}$ (or any linear combination 
of them) yield the minimum SNR while $Y^{(3)}$ yields maximum SNR. The remaining case,  
when $\eta (f) = 1$ all the $Y^{(I)}$ have the same SNR. 
Since the extremum values of SNR are only attained by the eigenvectors, the corresponding SNR curves constitute 
the bounding curves for any linear combination of $Y^{(I)}$ s. So our results determine 
the limiting sensitivities of data combinations cancelling laser frequency noise and 
optical bench motion noise. It is interesting to note that at the frequencies where the 
bounding curves intersect,  
{\it all} the data combinations belonging to the module have the same SNR.  

\begin{figure}
\begin{center}
\caption{\label{cap:etaplot}Plot of parameter $\eta $ as a function of $f$.
The points $\eta =1$ correspond to the frequencies at which all
the data combinations have the same SNR. $\eta >$1 corresponds to
the region in which data combination $Y^{(1)}$and $Y^{(2)}$ are
more sensitive than $Y^{(3)}$ and vice versa. }

\includegraphics{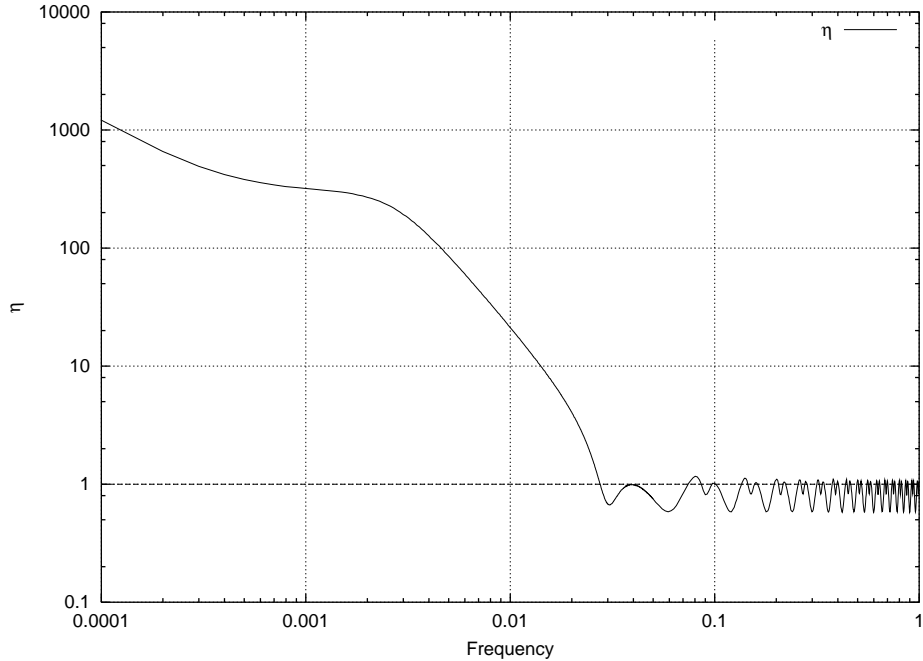}
\end{center}
\end{figure}

In the lower frequency range ($f \lsim 15\, mHz$), the combination $Y^{(3)}$(same
as that of the combination $\zeta $) has a very low sensitivity to
the gravitational wave signal and the generators $Y^{(1)}$ (or $Y^{(2)}$) has maximum
sensitivity to the signal. However, at high frequencies, the sensitivity curves of 
$Y^{(1)}$ and $Y^{(3)}$ intersect at several frequencies eg. $\sim 27$
mHz, $39$ mHz etc. While computing $\mathcal{H}^{(IJ)}$ and the average sensitivity, 
we assume a uniform
distribution of sources over polarisations and source directions in
the sky. Averaging over the polarisations is performed analytically
and the averaging over source directions performed using the Monte Carlo
method. The sensitivity, $S$ is defined following the reference \cite{LISArep},
$S=5\, \sqrt{\frac{B}{SNR}}$, here, $B=\frac{1}{T}$ and $T$ is
the observation time which we take to be one year. In the Figure \ref{cap:Plotsens},
we show the plots of $S$ for the basis elements $Y^{(I)}$, for comparison,
we also plot the sensitivity for the Michelson combination $X$. In the
Figure \ref{cap:etaplot}, we plot the ratio $\eta$ as a function of the frequency 
$f$. The points at which $\eta $ intersects the line $y=1$ corresponds to the points
where all the data combinations in the module have the same value of SNR.

\subsection{Operating LISA in a network mode}

In the preceding sections, we have shown that either $Y^{(1)}, Y^{(2)}$ or $Y^{(3)}$ 
maximise the SNR and they are orthogonal {\it i.e.} they are independent random variables. 
The sensitivity of LISA can be further improved
because, each of these generators can be realized as independent gravitational
wave detectors. Here we obtain the network SNR by taking $Y^{(I)}$
as independent outputs of a network of three detectors. We assume that the underlying 
noise is Gaussian \cite{SDP99} and the $Y^{(I)}$ follow the standard normal 
distribution.

\begin{figure}
\begin{center}
\caption{Plots showing the relative improvements (ratios) of SNRs for the three cases:
(i) Network SNR over the Michelson data combination (solid line).
(ii) Network SNR over the maximum of Max $[Y^{(1),}Y^{(3)}]$ (dotted line).
(iii) Max $[Y^{(1),}Y^{(3)}]$ over the Michelson (dashed line).
Here Max $[Y^{(1),}Y^{(3)}]$ is the maximum of the SNR of $Y^{(1)}$ and $Y^{(3)}$ 
over the bandwidth of LISA.}

\includegraphics{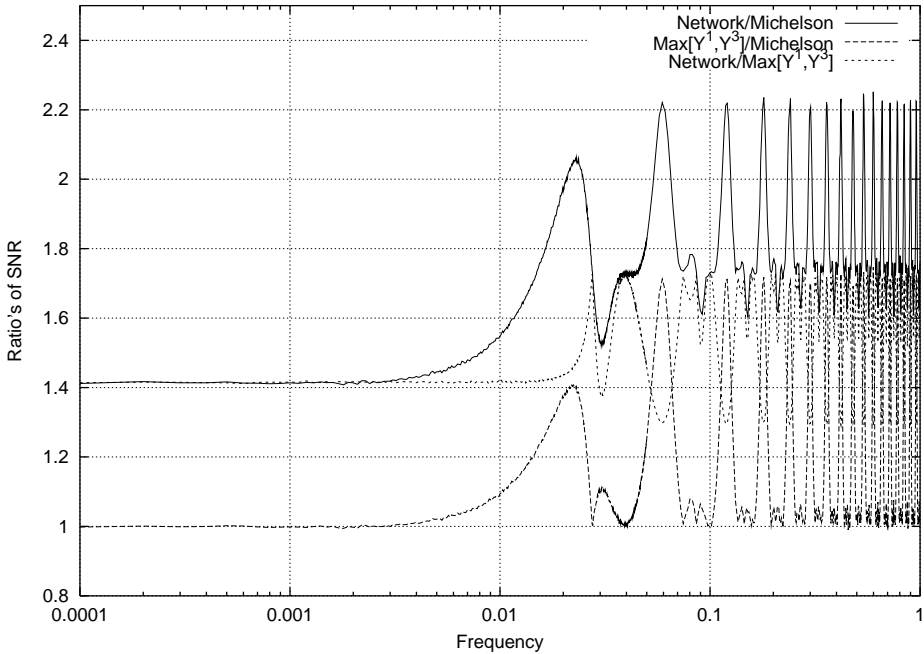}
\label{network}
\end{center}
\end{figure}

 As shown in \cite{SDP99}, if the noise of the individual detectors is uncorrelated
then the network likelihood ratio is just the product of individual likelihood ratios; 
the log network likelihood ratio is the sum of the individual log likelihood ratios.
Moreover, if the noise in the individual detectors is Gaussian, the surrogate statistic 
of the network, yields the network SNR as,
\begin{equation}
SNR_{network}^{2}\, \, =\, \sum_{I = 1}^{3} SNR_{(I)}^2 =\, 2\frac{h_{1}}{n_{1}}+\frac{h_{3}}{n_{3}}\, .
\label{eq:netSNR}
\end{equation}
The corresponding sensitivities are shown in the Figure \ref{cap:Plotsens}.
At low frequencies $f \lsim 15$ mHz, the improvement of the network SNR over the 
maximum of $Y^{(I)}$ is slightly greater than $\sqrt{2}$. This is because at low 
frequencies the data combination $Y^{(3)}$ is not very sensitive in comparison
with $Y^{(1)}$. The best improvement in the relative SNR is achieved at frequencies 
where all the data combinations are equally sensitive, that is, when  $\eta =1$. 
A gain of a factor of $\sqrt{3}$ is achieved at these points. 
In the Figure \ref{network}, we have plotted the relative improvements 
in the network SNR with respect to the Michelson combination and 
the maximum of $Y^{(1)}$ and $Y^{(3)}$.%

\section{SNR maximisation for a toy model of the LISA triangle rotating in a plane}

The LISA configuration consists of three spacecraft which orbit with 
complex trajectories in order to maintain a stable triangular configuration in 
an heliocentric orbit \cite{LISArep}. We note that the transfer functions of the beams $U^i, V^i$ are
computed in the frame of LISA. Hence, a given GW source which is
fixed on the celestial sphere will appear to follow a complex trajectory 
in this frame. This motion is the superposition of two rotations. One is just a 
rotation of the LISA triangle in its own plane. The other is its motion around the 
Sun in which the configuration follows the earth in an earthlike orbit. 
The period of the two rotations is identical and is one year in duration. Since 
LISA has a non-uniform directional response, it is a non-trivial problem to track 
the apparent motion of a `fixed' GW source in the LISA frame and then compute its SNR.

\begin{figure}
\begin{center}
\caption{\label{cap:3dplots}Plot of $\log \, S$, of the $Y^{(1)}$ and $Y^{(2)}$,
are displayed in (a) and (b) respectively, as a function of $\theta $
and $\phi $ for $f=1\, mHz$ and SNR=5 over a observation period
of one year. }

\includegraphics[  width=16cm]{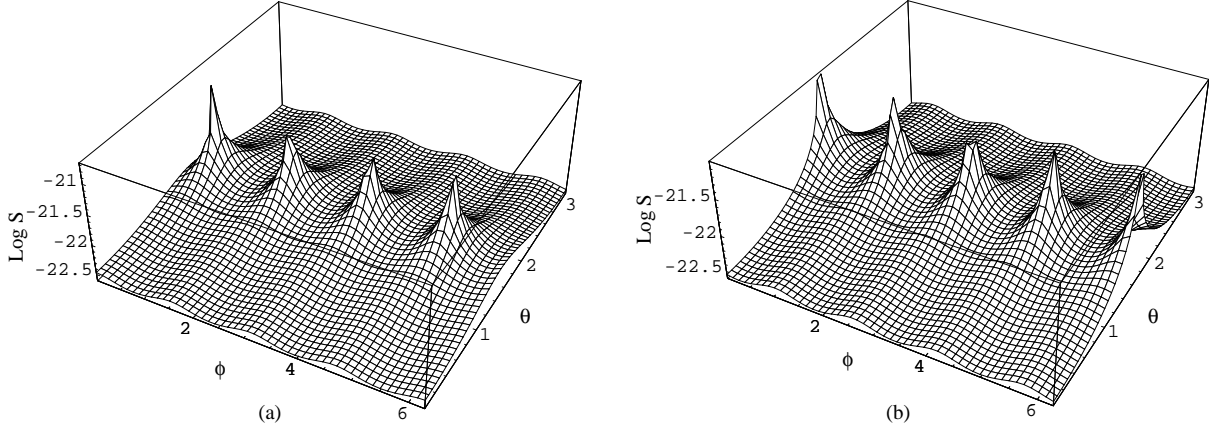}
\end{center}
\end{figure}

\begin{figure}
\begin{center}
\caption{\label{cap:phi-plot}Plot of sensitivity as a function angle $\phi$
at $\theta =\frac{\pi }{2}$ and $f=1\, mHz$ after averaging over
the polarisations.}

\includegraphics{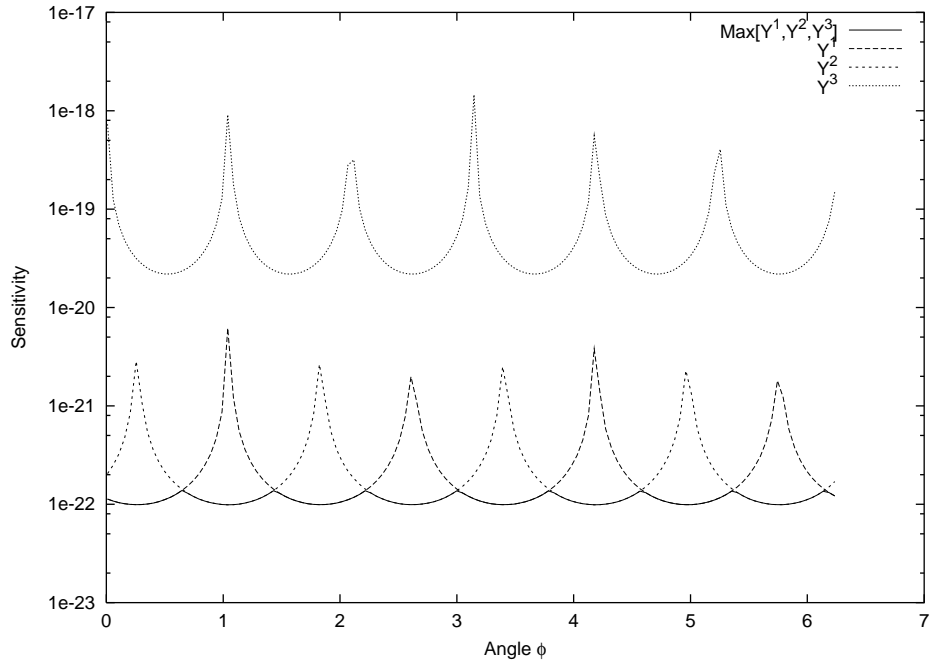}
\end{center}
\end{figure}

 We consider a toy model in which only one motion is taken into account, which is
the rotation of LISA about an axis orthogonal to the plane containing the three 
space-craft. We ignore the other motion of LISA around the Sun for simplicity. 
Following paper I, we choose the $X-Y$ plane to coincide with the plane formed by 
the three space-craft constituting the LISA triangle. The origin is chosen at the 
centre of the equilateral triangle (for computing the response, we take all the arms 
to be of equal length). The positive $X$-axis passes through space-craft 2. 
Figure 1 shows how the $X-Y$ axes have been chosen. The $Z$-axis is orthogonal to 
this plane with the positive direction given by the right-handed convention.
Under these assumptions, the source would appear to move in the sky in a circular 
trajectory about the $Z$-axis with a period of one year. In terms of polar coordinates 
$(\theta, \phi)$, the motion of such a source is uniform along $\phi$ with a constant 
value of $\theta$. If one tracks a source with a single data combination, say 
$Y^{(1)}$, because the directional sensitivity of $Y^{(1)}$ is non-uniform, it does not 
track the source optimally. Figure (\ref{cap:3dplots}) displays the  
3-dimensional sensitivity plots of $Y^{(1)}$ and $Y^{(2)}$ as a function of the 
angles $(\theta, \phi)$ for a monochromatic source at the GW frequency of $1$ mHz and 
the signal uniformly averaged over the polarisations (this average is computed 
analytically). It is obvious from these plots that the sensitivity is a highly 
non-uniform function of source direction. We consider here the case when  
the source with GW frequency $1$ mHz lies in the plane of the LISA triangle, that is, 
the GW source lies in the $\theta =\frac{\pi }{2}$ plane. We choose this value because,  
in the plane of LISA, the variations in the sensitivity are maximum
for the combinations $Y^{(1)}$ and $Y^{(2)}$. 
From the Figures (\ref{cap:3dplots}) and (\ref{cap:phi-plot}), we note that the 
generators  $Y^{(I)}$ have zero sensitivity at few values of $\phi$. This implies that
no single data combination, even if it is a linear combination of these, can give optimal 
sensitivity. To obtain best results, one needs to maximise the SNR over the linear 
combinations $\alpha_{(I)} Y^{(I)}$.

We analyse two strategies for optimising the sensitivity: 
\begin{description}

\item [(a)] We consider the cyclic permutations of $Y^{(1)}$ and compute the maximum 
sensitivity using these three data combinations. 
 
\item [(b)] We take the maximum of the linear combinations $\alpha_{(I)} Y^{(I)}$ 
where $\alpha_{(I)}$ are complex numbers. At each value of $\phi$, a different 
linear combination of $Y^{(I)}$ is optimal. Thus the $\alpha_{(I)}$ for which the 
SNR is maximised are, in general, functions of $\phi$. 

\end{description}

\begin{figure}
\begin{center}
\caption{\label{cap:alpha_plot} The coefficients $\alpha _{(I)}$ maximising 
the SNR of the combination $\alpha _{(I)}Y^{(I)}$ are plotted as functions of
$\phi $ for the frequency $1$ mHz and $\theta = \pi/2$.}

\includegraphics{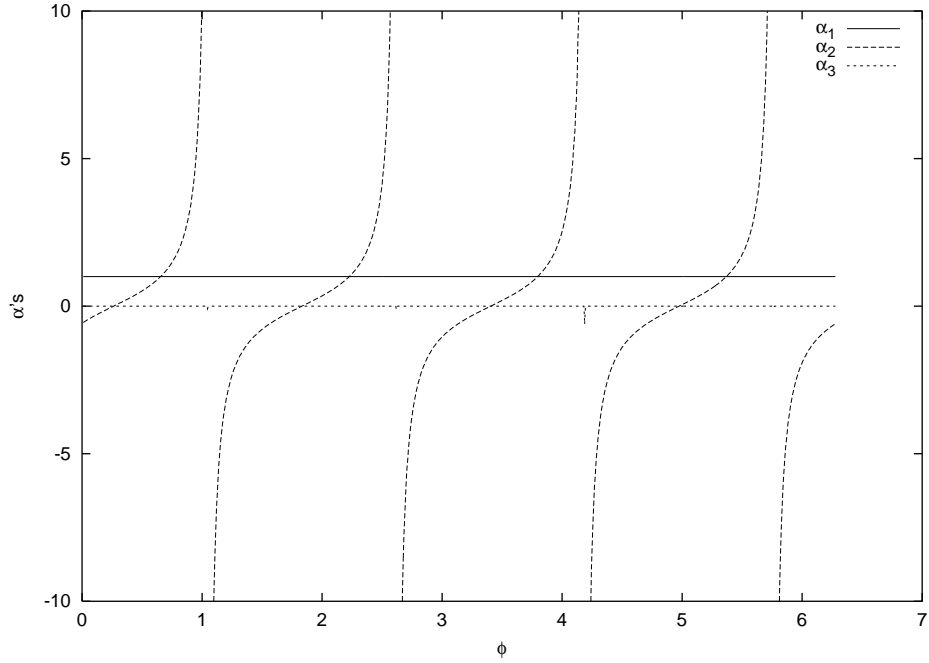}
\end{center}
\end{figure}

\begin{figure}
\begin{center}
\caption{\label{cap:Opti_phi} Sensitivity curves for the toy model of LISA are given 
as functions of  $\phi $ at $f=1\, mHz$. The curves represented by broken lines 
correspond to the $Y^{(1)}$ and its cyclic permutations. The solid line curves 
represent the sensitivities for the two strategies: (a) taking the maximum of 
$Y^{(1)}$ and its cyclic permutations (thin solid line); (b) taking maximum 
over the linear combinations of $Y^{(I)}, I = 1, 2, 3$ (thick solid line).}

\includegraphics{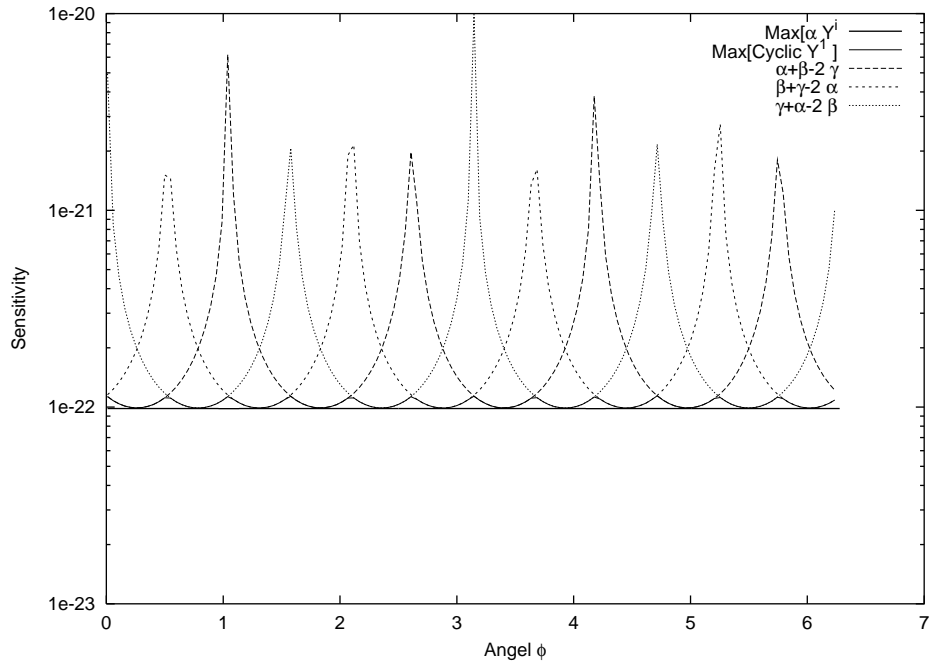}
\end{center}
\end{figure}

The results of these analyses are described in Figure (\ref{cap:Opti_phi}). 
For the strategy (a), the improvement in sensitivity averaged over $\phi$ is $\sim$ 49 \%;
while for the strategy (b), the improvement in sensitivity averaged over 
$\phi$ is $\sim$ 59 \%. In addition Figure (\ref{cap:alpha_plot}) shows the plots of 
$\alpha_{(I)}$ as function of $\phi$, which maximise the SNR. 

The strategy (a) does not give the best SNR as it maximises the response over only a 
set of three data combinations. On the other hand, strategy (b) maximises the SNR over 
the module and is therefore superior. It is interesting to note that the maximised 
SNR is constant as a function of $\phi$ and is the same as the SNR of $Y^{(1)}$ 
maximised over $\phi$. We remark that the strategy (b) gives only marginal improvement 
over strategy (a), and also strategy (a) is easier to implement requiring relatively 
fewer computations.

\begin{figure}
\begin{center}
\caption{\label{cap:toysens}Plot of sensitivity $S$ as a function of frequency,
for the toy model.}

\includegraphics{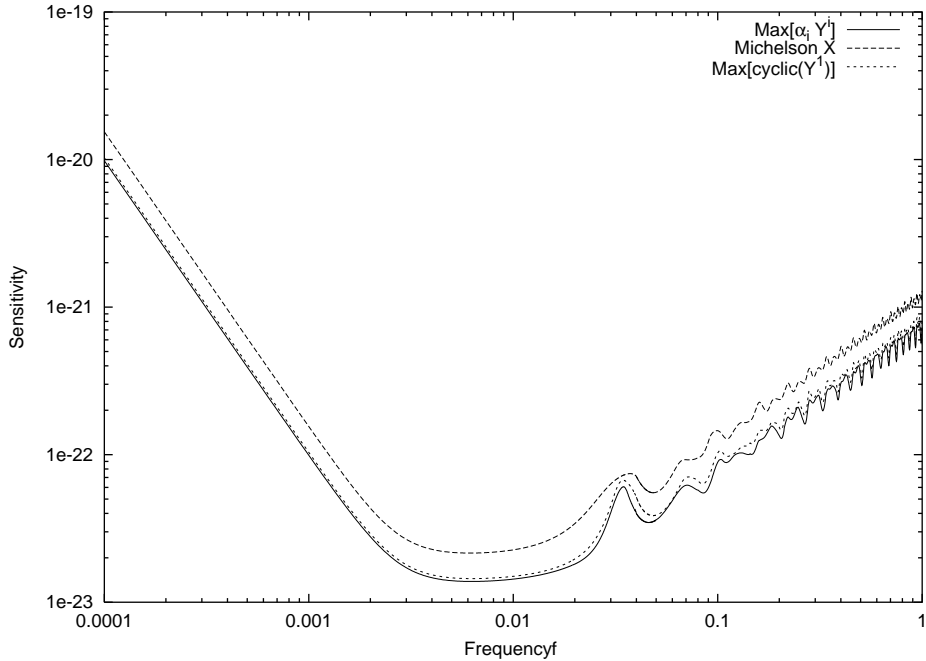}
\end{center}
\end{figure}

As seen for a GW source with a frequency of $1$ mHz, strategy (b) 
gives a constant value for the maximum sensitivity over $\phi$ 
(see Figure (\ref{cap:Opti_phi})). 
Thus the maximum sensitivity is independent of $\phi$ for a given frequency. 
We further extend this analysis to other frequencies over the band-width of LISA,  
and obtain analogous results to the case of 1 mHz. This maximum 
sensitivity however depends on the frequency.
In Figure (\ref{cap:toysens}), we present the maximum sensitivity curves as
functions of frequency. For comparison, we perform analogous computations for the 
Michelson data combination $X$ and the maximum of the cyclic permutations of 
$Y^{(1)}$. In all these cases, the GW amplitude is first averaged over the polarisations 
 and then for a fixed frequency, the sensitivity as a function of $\phi$ is 
computed and then averaged over $\phi$. Finally, this exercise is carried out for 
frequencies across the LISA band-width. 

The 1 mHz case is representative of the low frequency regime $\lsim$ 3 mHz where the 
improvement in sensitivity is about 49 \% for strategy (a) and 59 \% for strategy (b). 
These improvements scale up at higher frequencies at say, $\sim$ 15 mHz, 
to 56 \% and 67 \% , respectively, for the two strategies (a) and (b).
 
\section{conclusion}

In this paper, we have employed our previously set up formalism from paper I for two 
important applications: 

The first application is the extremisation of SNR. 
We have considered binaries as GW sources whose frequencies 
change at most adiabatically (monochromatic sources are included), that is, even if 
the frequency changes during the observation time, the change in SNRs under 
consideration is insignificant. 
The signal covariance matrix has been computed for which the signal is averaged 
over polarisations and directions. We have then computed the noise covariance matrix 
for LISA and obtained its eigenvectors and eigenvalues. This 
matrix has the same eigenvectors, resulting in calculational simplification. We have 
shown that the SNR for any data combination in the module, lies between  
the upper and lower bounds which are determined by the eigenvectors of both matrices. 
We have further shown that the bounding SNR curves of the 
eigenvectors have multiple intersections within the band-width of LISA - $10^{-4}$ - 1 Hz. 
We have obtained the following results for the improvement in SNR:
The improvement of SNR of the upper-bound over the Michelson combination goes upto 70 \% , 
at high frequencies $\gsim 5$ mHz, however, at low frequencies $\lsim$ 5 mHz, both 
have the same sensitivity. Since the eigenvectors are 
independent random variables, a `network' SNR  of independent detectors \cite{SDP99} 
has been constructed from likelihood 
considerations which gives an improvement between $\sqrt {2}$ and $\sqrt {3}$ over the 
maximum of SNRs of the eigenvectors. The improvement of the network SNR over the 
Michelson combination is about 40 \% at low frequencies $\lsim 3$ mHz and rises above 
100 \% at high frequencies. 
\par
The second application is a simple toy model of LISA rotating in its own plane. 
For this model we estimated the improvement of SNR by optimally switching  
from one data combination to another. We have shown that, if this strategy is used, 
it is possible to improve the SNR by about 55 \% on an average over the band-width of 
LISA. We also show that if we instead maximise over the module, the SNR improves 
by about 60 \% on an average over the LISA band-width. These improvements are obtained 
using just one data combination, namely, the eigenvector $Y^{(1)}$. The 
SNR improvement in both cases is larger at higher frequencies $\gsim 10$ mHz than at 
low frequencies. 

\ack
The authors would like to thank B. F. Schutz for in depth discussions which provided 
valuable physical insights and for suggesting the toy model
scenario. The authors would like to thank T. Prince and M. Tinto for
useful discussions.

\end{document}